\documentclass[journal,twoside,web]{ieeecolor}
\usepackage{generic}
\usepackage{cite}
\usepackage{amsmath,amssymb,amsfonts}
\usepackage{algorithmic}
\usepackage{graphicx}
\usepackage{textcomp}
\usepackage{hyperref}
\usepackage{siunitx}
\def\BibTeX{{\rm B\kern-.05em{\sc i\kern-.025em b}\kern-.08em
    T\kern-.1667em\lower.7ex\hbox{E}\kern-.125emX}}
\begin{document}

\title{\LARGE Pseudo-random sequences for low-cost operando impedance measurements of Li-ion batteries}

\author{Jussi Sihvo, Noël Hallemans, Ai Hui Tan, David A.\ Howey, Stephen. R.\ Duncan, Tomi Roinila 
\thanks{This paper has been originally submitted for review on 5th May 2025 to IEEE.}
\thanks{Jussi Sihvo is with the Department of Energy, Aalborg University, Aalborg, Denmark (e-mail: jmesi@energy.aau.dk). }
\thanks{Noël Hallemans, David Howey, and Stephen Duncan are with the Department of Engineering Science, University of Oxford, UK (e-mail: noel.hallemans@eng.ox.ac.uk, david.howey@eng.ox.ac.uk, stephen.duncan@eng.ox.ac.uk). }
\thanks{Ai Hui Tan is with the Faculty of Artificial Intelligence and Engineering, Multimedia University, Cyberjaya, Malaysia (e-mail: htai@mmu.edu.my).}
\thanks{Tomi Roinila is with the Department of Electrical Engineering, Tampere University, Tampere, Finland (e-mail: tomi.roinila@tuni.fi).}
}

\newcommand{\NH}[1]{{\color{blue}#1}}

\maketitle

\begin{abstract}
Operando impedance measurements are promising for monitoring batteries in the field. In this work, we present pseudo-random sequences for low-cost operando battery impedance measurements. The quadratic-residue ternary sequence and direct-synthesis ternary sequence exhibit specific properties related to eigenvectors of the discrete Fourier transform matrix that allow computationally efficient compensation for drifts and transients in operando impedance measurements. We describe the application of pseudo-random sequences and provide the data processing required to suppress drift and transients, validated on simulations. Finally, we perform experimental operando impedance measurements on a Li-ion battery cell during fast-charging, demonstrating the applicability of the proposed method. It's low-cost hardware requirements, fast measurements, and simple data-processing make the method practical for embedding in battery management systems.
\end{abstract}


\section{Introduction}
\IEEEPARstart{L}{i-ion} batteries in electric vehicles and electronic devices require active diagnostics. This is done by battery management systems (BMS) that monitor different states (state-of-health, state-of-charge (SOC), temperature, etc.) to optimise battery management and fast-charging protocols, directly affecting the safety, efficiency, and lifetime of the battery and application. The BMS typically estimates the states from measured current, voltage, and temperature data. Along with these measurements, the impedance is valuable for state monitoring \cite{Meddings2020,jones2022impedance,plett2023battery}---mainly due to the separation of time scales corresponding to different physical processes in the battery.

However, impedance data is not widely used in BMS applications. One reason for this is the high implementation cost of conventional electrochemical impedance spectroscopy (EIS) \cite{wang2021electrochemical}, requiring extensive hardware for careful measurements \cite{Howey2014}, and for generation of sinusoidal excitation. Moreover, conventional EIS measurements can only be performed in steady-state, which is a general requirement in many system identification applications \cite{ljungSystemidentification,Pintelon2012} but rarely occurs in practice for batteries in the field \cite{hallemans2023electrochemical}. In the latter situation, impedance must be measured \textit{in operando}---during normal operation of the battery in its host application (e.g.~while charging or driving an electric vehicle). However, charging and discharging of the battery can change the local conditions, such as SOC and temperature, which makes the impedance time-varying during experiments \cite{hallemans2023electrochemical}. In addition, the battery is never at steady-state, as the charging/discharging current creates voltage drifts, and transient effects are present \cite{hallemans2022operando}. All of these have a corrupting effect on the \textit{operando}-measured impedance spectrum, hindering further analysis compared to steady-state conditions.

Existing \textit{operando} EIS methods are based on sinusoidal \cite{Huang2015,Stoynov1993} or multisine (sum-of-sine) \cite{breugelmans2012odd,koster2017dynamic,hallemans2022operando} excitations. In particular, multisines inject all frequencies simultaneously, which reduces measurement time and minimises the impact of non-stationary conditions compared with conventional EIS \cite{hallemans2023electrochemical}. These methods heavily focus on suppressing voltage drifts and modeling time-variation with complex techniques such as the four-dimensional approach \cite{Stoynov1993}, wavelet transforms \cite{Li2019}, or frequency-domain modelling \cite{hallemans2022operando}. Despite the potential and good performance of these methods, the use of multisine (or sine) excitations makes them difficult to apply with low-cost onboard electronics. Moreover, the memory and computational requirements for data processing exceed the majority of BMS processors. Impedance measurement techniques that are technically and economically viable for use in \textit{operando} need to be developed for real-world applications.

An interesting alternative to multisines is provided by pseudo-random sequences (PRS) \cite{Godfrey1993,Tan2013}, which have been widely used for identifying electrical motors \cite{Vermeulen2002}, batteries \cite{Sihvo2020a,Sihvo2025}, fuel cells \cite{Mahlangu2022}, and power-distribution systems \cite{Roinila2021}. They have several advantages over other excitations: their broadband signal characteristics allow injection of many frequencies simultaneously (minimising measurement time), the sequences typically only have two or three levels (making them relatively easy and cheap to apply in practice), and spectral leakage can be avoided due to periodicity \cite{Pintelon2012}. Moreover, the signal spectrum can be designed to excite a specific set of harmonics, which is useful for identification of nonlinear systems. These properties make PRSs particularly suitable for low-cost impedance measurements for BMS applications. 

While reduced hardware costs make PRS methods favorable over multisines for \textit{operando} impedance measurements, the heavy data processing required for drift and transient suppression is still a challenge for BMS applications. However, some ternary (three-level) pseudo-random sequences exhibit specific properties that enable very simple suppression of voltage drifts and transient effects. These properties are related to the sequences having three levels and being eigenvectors of the discrete Fourier transform (DFT) matrix \cite{McClellan1971}. Examples of signals with these properties are the quadratic-residue ternary (QRT) sequence \cite{Godfrey1999} and the direct-synthesis ternary (DST) sequence, which has suppressed harmonics for identifying systems exhibiting strong nonlinear distortions \cite{Tan2013}. The DST sequence has recently been demonstrated for battery impedance measurements under dynamic current conditions \cite{Sihvo2025}. However, the full theoretical derivation and validation of the DFT eigenvector properties, and drift and transient suppression have not yet been presented.

This paper provides a guide on using QRT and DST sequences for \textit{operando} battery impedance measurements. The theory behind the eigenvector properties is described and illustrative sequences are shown. We explain how these can be implemented in practice and estimate impedance during operation, while suppressing voltage drifts and transients. The proposed method is validated through simulations and experiments from charging of a Li-ion cell. The properties of the presented PRSs enable extraction of impedance during operation within a single excitation period and without complex correction methods. As the methods provide fast measurements with low implementation complexity, they are a pragmatic and economically feasible solution for \textit{operando} impedance measurements in modern high-performance BMS algorithms.

\section{Ternary sequences with DFT eigenvector properties} 
\label{PRS sequences for non-stationary system identification}

\subsection{The quadratic-residue ternary (QRT) sequence}
The QRT sequence---originating from the Legendre symbol introduced in 1798---is a discrete periodic real-valued ternary sequence of length $N_\text{QRT}$ defined as \cite{Narasimha1976,Godfrey1999}
\begin{align}
    u_\text{QRT}(n) &= 
\begin{cases} 
0 & \text{for $n=0$}\\
	1 &	\text{for $n\in \mathbb{S}$}\\
	-1 & \text{else},
 \end{cases}
 \label{eq:timeQRT}
\end{align}
with $n = 0,1,\hdots,N_\text{QRT}-1$,
\begin{align}
    \mathbb{S} = \{1^2, 2^2,\hdots, (N_\text{QRT}-1)^2\}_\text{mod $N_\text{QRT}$},
\label{eq:setSQRT}
\end{align}
and $N_\text{QRT}$ an odd prime satisfying $N_\text{QRT}= 4z \pm 1$ with $z\in \mathbb{N}$. We can analyse this sequence in the frequency domain using the normalised DFT given as
\begin{align}
   U_\text{QRT}(k)= \frac{1}{\sqrt{N_\text{QRT}}}\sum_{n=0}^{N_\text{QRT}-1} u_\text{QRT}(n)e^{-\frac{j2\pi kn}{N_\text{QRT}}},
    \label{eq:DFT}
\end{align}
with $k=0,1,...,N_\text{QRT}-1$ the harmonic index. Unlike most PRSs, the QRT sequence is an eigenvector of the DFT matrix;
\begin{align}
         \mathbf{U}_\text{QRT} = \mathbf{F}\mathbf{u}_\text{QRT} = \lambda \mathbf{u}_\text{QRT},
     \label{eq6}
\end{align}
with
\begin{align}
    \mathbf{u}_\text{QRT}=[u_\text{QRT}(0)\  u_\text{QRT}(1)\ \hdots\ u_\text{QRT}(N_\text{QRT}-1)]^\top,
\end{align} 
\begin{align}
    &\mathbf{F} = \frac{1}{\sqrt{N_\text{QRT}}}
  \begin{bmatrix} 
  1 & 1 & 1 &  \dots & 1\\
  1 & w^1 & w^2 & \dots & w^{(N_\text{QRT}-1)} \\
  1 & w^2 & w^4 & \dots & w^{2(N_\text{QRT}-1)}\\
  \vdots & \vdots & \vdots & \ddots & \vdots \\
  1 & w^{N_\text{QRT}-1} & w^{2(N_\text{QRT}-1)} &\dots &  w^{(N_\text{QRT}-1)^{2}}
  \end{bmatrix}, 
  \label{eq7}
\end{align}
and $w=e^{-\frac{j2\pi}{N_\text{QRT}}}$. For a given vector $\mathbf{u}_\text{QRT}$, the associated eigenvalue $\lambda$ takes \textit{one} out of four (either purely real or purely imaginary) values, depending on the length and symmetry properties of the eigenvector \cite{Narasimha1976},
\begin{align}
\lambda \in \{1,-1,j,-j\}.
    \label{eq:lambda}
\end{align}
This effect is related to the DFT eigenvalue multiplicity problem presented in \cite{McClellan1971}. The time- and frequency-domain values of the QRT sequence are, hence, correlated,
\begin{align}
U_\text{QRT}(k) = \lambda u_\text{QRT}(k)\qquad  k=0,1,\hdots,N_\text{QRT} - 1.
\label{eq:eigenvectorentry}
\end{align}
  
As $u_\text{QRT}(n)$ \eqref{eq:timeQRT} only takes two non-zero values ($-1$ and $1$), it follows from \eqref{eq:eigenvectorentry} that $U_\text{QRT}(k)$ also takes two non-zero values ($-\lambda$ and $\lambda$). We now introduce two sets containing the indices at which the non-zero values take place,
\begin{align}
\begin{split}
    \mathbb{K}_+&=\{k\  \vert \ u_\text{QRT}(k)=1\}\\
\mathbb{K}_-&=\{k\  \vert \ u_\text{QRT}(k)=-1\},
\label{eq:IposIneg}
\end{split}
\end{align}
and find that 
\begin{align}
     U_\text{QRT}(k) &= \begin{cases}
     0 & \text{for } k=0\\
      \lambda  & \text{for } k\in \mathbb{K}_+\\
      -\lambda  &  \text{for } k\in \mathbb{K}_-.
  \end{cases} 
  \label{eq:UQRTcases}
\end{align}
The two non-zero values of $U_\text{QRT}(k)$, hence, take place on two sets of intertwined indices, $\mathbb{K}_+$ and $\mathbb{K}_-$ \eqref{eq:IposIneg}. Moreover, the non-zero values of $U_\text{QRT}(k)$ \eqref{eq:UQRTcases} are complements of each other (either $\lambda$ or $-\lambda$). Hence, from \eqref{eq:UQRTcases} we write that
\begin{align}
   U_\text{QRT}(k_+) = -U_\text{QRT}(k_-) \ \ \forall\  k_+\in \mathbb{K}_+ \text{ and } k_-\in \mathbb{K}_-.
   \label{eq:kposknegminusQRT} 
\end{align}
We will exploit this property for suppressing drift and transient signals in \textit{operando} impedance measurements (Section~\ref{Operando impedance measurements}).

\begin{figure}
\begin{center}
\includegraphics[width=\linewidth]{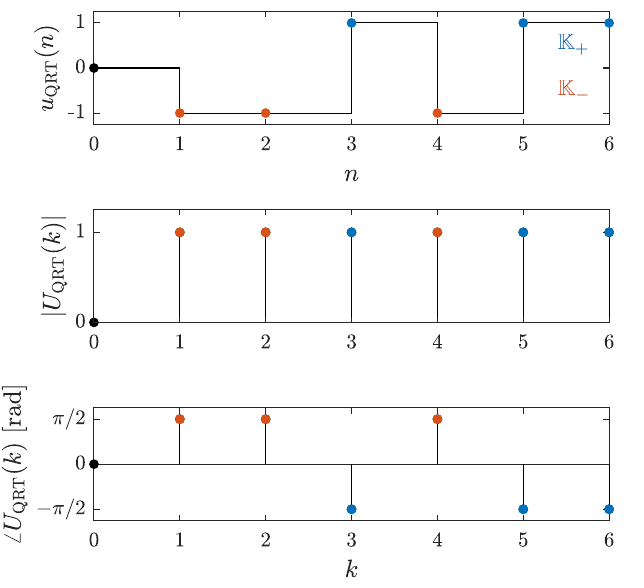} 
\caption{Time- and frequency-domain characteristics of QRT sequence of length 7. Top: time-domain signal. Middle: DFT magnitude response. Bottom: Phase response. The values at the sets $\mathbb{K}_+=\{3,5,6\}$ and $\mathbb{K}_-=\{1,2,4\}$ are, respectively, indicated in blue and red.}
\label{fig:QRT}
\end{center}
\end{figure}

An example QRT sequence of length $N_\text{QRT}=7$ with eigenvalue $\lambda = -j$ is shown in Fig.~\ref{fig:QRT}. The QRT excites all harmonics (except DC) with uniform magnitudes of one, which is common for many types of PRSs used for system identification \cite{Tan2002}. The values $U_\text{QRT}(k)$ at the harmonics $\mathbb{K}_+$ (blue, time-domain values of 1) have a phase difference of $\pi$ compared to the values at harmonics $\mathbb{K}_-$  (red, time-domain values of $-1$). 

\subsection{The direct-synthesis ternary (DST) sequence}
The QRT sequence defined above excites all harmonics. However, when identifying nonlinear systems, it is useful to suppress certain harmonics. Suppressing even harmonics allows detection and classification of even and odd nonlinearities and estimation of the best linear approximation of the system \cite{schoukens2005identification,schoukens2016linear}. In this section we extend the QRT to suppress second- and third-order harmonics while still satisfying the eigenvector property at excited harmonics, resulting in the DST sequence \cite{Tan2013}.

The DST sequence is generated by combining two sequences; a special and a basic sequence. The special sequence of fixed length of six yields
\begin{align}
    \mathbf{u}_\text{special} = [0 \hspace{2mm}  -1 \hspace{2mm} -1 \hspace{2mm} 0 \hspace{2mm} 1 \hspace{2mm} 1] 
    \label{eq:uspecial}
\end{align}
and the basic sequence must be a suitable PRS (e.g.\ the QRT sequence---see list in \cite{Tan2013}) with length
\begin{align}
    N_\text{basic} \in \left\{5 + 6p, 7+6p \hspace{1mm} | \hspace{1mm} p \in \mathbb{N}\right\}.
    \label{eq:NDST}
\end{align}
Subsequences $\mathbf{a}$ and $\mathbf{b}$ are then formed by concatenating the special and basic sequence, respectively, $N_\text{basic}$ and six times ($N_\text{special}=6$), resulting both in a length $N_\text{DST}=6N_\text{basic}$,
\begin{align} 
		\mathbf{a} &= [\overbrace{\mathbf{u}_\text{special}\ \mathbf{u}_\text{special}\  \hdots\  \mathbf{u}_\text{special}}^{N_\text{basic} \text{ times}}]&  &\in \mathbb{Z}^{1\times 6N_\text{basic}}\nonumber\\
		\mathbf{b} &= [\underbrace{\mathbf{u}_\text{basic}\  \mathbf{u}_\text{basic} \ ... \ \mathbf{u}_\text{basic}}_\text{6 times}]& &\in \mathbb{Z}^{1\times 6N_\text{basic}}.
\label{eq:ab}
\end{align}
The DST sequence is finally obtained as the element-wise multiplication of $\mathbf{a}$ and $\mathbf{b}$,
\begin{align}
     u_\text{DST}(n) = a(n) b(n)\qquad n=0,1,\hdots,N_\text{DST}-1.
     \label{eq:uDSTn} 
\end{align}
The procedure above is designed to suppress second- and third-order harmonics in the DFT $U_\text{DST}(k)$, while keeping the DFT eigenvector property \eqref{eq:eigenvectorentry} at the excited harmonics. Indeed, when taking a QRT sequence $u_\text{QRT}(n)$ with length $N_\text{QRT}$ and eigenvalue $\lambda_\text{QRT}$ as basic sequence, it holds that the DFT of the DST sequence is
\begin{align}
    U_\text{DST}(k)=
    \begin{cases}
        \lambda_\text{DST} u_\text{DST}(k)& \text{for } k\in \mathbb{K}\\
        0 & \text{else},
    \end{cases}
    \label{eq:Theorem1Eigenvector}
\end{align}
with $\lambda_\text{DST}=j\sqrt{2}r\lambda_\text{QRT}$, $r\in\{-1,1\}$ depending on the QRT sequence, and excited harmonics
\begin{align}
      \mathbb{K}=&\{1+6p,5+6p \ \vert \ p =0,1,\hdots,N_\text{QRT}-1\}\nonumber\\
      &\setminus \{N_\text{QRT}, 5N_\text{QRT}\}.
      \label{eq:Theorem1_I}
\end{align}
The proof is given in the Appendix. Using \eqref{eq:Theorem1Eigenvector}, we find that the spectrum of the DST sequence yields
\begin{align}  
      U_\text{DST}(k) = &\begin{cases}
      \lambda_\text{DST}   &\text{for } k\in \mathbb{K}_+\\
      -\lambda_\text{DST}  &\text{for } k\in \mathbb{K}_-\\
      0  &\text{else},
  \end{cases}
\label{eq:UDSTk}
\end{align}
with
\begin{align}
    \mathbb{K}_+&=\mathbb{K}\cap\{k\  \vert \ u_\text{DST}(k)=1\}\nonumber\\
    \mathbb{K}_-&=\mathbb{K}\cap\{k\  \vert \ u_\text{DST}(k)=-1\}.
\end{align}
Similar to \eqref{eq:kposknegminusQRT} for the QRT sequence, the DST sequence also has the property that
\begin{align}
   U_\text{DST}(k_+) = -U_\text{DST}(k_-) \ \ \forall\  k_+\in \mathbb{K}_+ \text{ and } k_-\in \mathbb{K}_-,
   \label{eq:kposknegminusDST} 
\end{align}
which we will exploit for suppressing drifts and transients in \textit{operando} impedance measurements.
\begin{figure}
\begin{center}
\includegraphics[width=\linewidth]{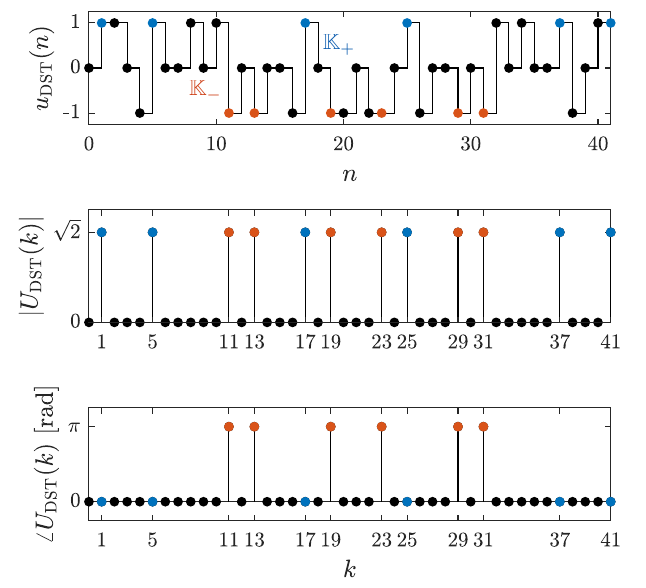} 
\caption{Time- and frequency-domain characteristics of the DST sequence of length 42. Top: time-domain signal. Middle: DFT magnitude response. Bottom: Phase response. The values at the sets $\mathbb{K}_+=\{1,5,17,25,37,41\}$ and $\mathbb{K}_-=\{11,13,19,23,29,31\}$ are, respectively, indicated in blue and red.}
\label{fig:DST}
\end{center}
\end{figure}

Fig.~\ref{fig:DST} shows an example of a DST sequence with the QRT sequence of length $7$ as its basic sequence. This DST sequence has length $N_\text{DST}=42$. We note that the DST is still an eigenvector of the DFT matrix, but only at the indices $\mathbb{K}$ \eqref{eq:Theorem1_I} corresponding to the excited harmonics, which for $N_\text{DST}=42$ yields
\begin{align}
    \mathbb{K}=\{1,5,11,13,17,19,23,25,29,31,37,41\}.
\end{align}
\section{Practical measurements with PRS signals}\label{Practical measurements with PRS signals}
We now detail how to apply the QRT and DST sequences to real-life systems by making them continuous, and how to sample current and voltage data. 

\subsection*{Zero-order hold reconstruction}
Zero-order hold reconstruction is often used to apply a continuous excitation signal to a system under test \cite{Schoukens2018}. The applied current signal $i(t)$, with $t$ the continuous time, is mathematically obtained by convolving the discrete sequence $u(n)$ with a top-hat function of length $T_\text{zoh}$, resulting in the black staircase signals in Figs.~\ref{fig:QRT} and \ref{fig:DST}, and multiplying with the amplitude of the excitation $C$,
\begin{align}
    i(t)=C\ \text{zoh}_{T_\text{zoh}}\{u(n)\}.
\end{align}
As a result of the reconstruction, the spectrum of the applied continuous signal $i(t)$ is filtered,
\begin{align}
    I(\omega)=C\ \text{ZOH}(\omega)U(\omega),
    \label{eq:IZOHId}
\end{align}
with $U(\omega)$ the Fourier transform of the repeated sequence $u(n)$ with intervals $T_\text{zoh}$ and 
\begin{align}
    \text{ZOH}(\omega)=e^{-j\frac{\omega T_\text{zoh}}{2}}\mathrm{sinc}\Big(\frac{\omega T_\text{zoh}}{2}\Big).
\end{align}
This filtering operation can be visualised in the current spectrum $I(k)$ in Fig.~\ref{fig:steadyStateMeasurement}. Two important observations should be made:
\begin{itemize}
    \item The zero-order hold period $T_\text{zoh}$ sets the bandwidth of the impedance measurement.  The period of the excitation is $T_\text{p}=N_\text{p}T_\text{zoh}$, with $N_\text{p}$ the length of the QRT or DST sequence. The lowest excited frequency is hence $f_\text{min}=1/T_\text{p}$. Regarding the maximal frequency, $I(\omega)$ has a first zero-crossing at $\omega=2\pi f_\text{zoh}$, with $f_\text{zoh}=1/T_\text{zoh}$, and has further zero-crossings at all  integer multiples. As the magnitude of the excitation already decreases significantly at frequencies lower than $f_\text{zoh}$ (therefore decreasing the signal-to-noise ratio), we measure impedance up to $f_\text{max}=2f_\text{zoh}/3$. In our experiments we set $f_\text{zoh}=1.5$~kHz so that we can measure frequencies up to \SI{1}{kHz} (as in Fig.~\ref{fig:steadyStateMeasurement}). For the experiments we use a DST sequence of length $N_\text{p}=10002$, such that $T_p=10002/1500=$ \SI{6.668}{s}, and the lowest frequency $f_\text{min}=$ \SI{0.15}{Hz}. Note that  $f_\text{zoh}$ here is the same as the ``generating frequency'', a commonly used term in many related studies \cite{Vermeulen2002,Sihvo2020a,Sihvo2025,Mahlangu2022,Roinila2021}.
    \item Properties \eqref{eq:kposknegminusQRT} and \eqref{eq:kposknegminusDST} hold for the discrete sequences, however, they do not hold for the continuous spectrum $I(\omega)$. Instead, 
    \begin{align}
    I(\omega_{k_+}) = -\frac{\text{ZOH}(\omega_{k_+})}{\text{ZOH}(\omega_{k_-})}I(\omega_{k_-}) \ \ \forall  \ \begin{array}{l}  k_+\in \mathbb{K}_+ \\ 
    k_-\in \mathbb{K}_-,
    \end{array}
    \label{eq:ZOHfactor}
    \end{align}
with $\omega_k=2\pi k/T_\text{p}$. Still, for harmonics $k_+$ and $k_-$ that are close to each other, we have that $ I(\omega_{k_+}) \approx -I(\omega_{k_-})$.
\end{itemize}

\subsection*{Windowing}
To avoid spectral leakage, it is important to measure the applied current and voltage response for an integer number of periods $P\in \mathbb{N}$, that is, within a window $[0,T]$ with $T=PT_\text{p}$ and $T_\text{p}$ the period of the excitation signal \cite{Schoukens2018}. In this work we use one period ($P=1$).

\subsection*{Sampling}
The windowed applied current $i(t)$ and voltage response of the battery $v(t)$ are sampled at rate $f_\text{s}$ to obtain time series 
\begin{align}
\begin{split}
    \textbf{i}&=[i(0)\ i(T_\text{s})\ \hdots \ i((N-1)T_\text{s})]\\
    \textbf{v}&=[v(0)\ v(T_\text{s})\ \hdots \ v((N-1)T_\text{s})],
    \label{eq:sampledData}
\end{split}
\end{align}
with sampling period $T_\text{s}=1/f_\text{s}$ and number of samples $N=T/T_\text{s}=PN_\text{p}f_\text{s}/f_\text{zoh}$. 

The sampling frequency $f_\text{s}$ is typically chosen larger than twice the maximum frequency in the measured signal to avoid aliasing. However, the QRT and DST sequences are not band-limited, and, hence, aliasing can only be avoided by using an anti-aliasing filter at a frequency $f_\text{N}>f_\text{max}$ and sampling at $f_\text{s}>2f_\text{N}$. If no anti-aliasing filter is available in the measurement setup, aliasing cannot be avoided. However, one can suppress the effect of alias by oversampling. In our measurements, we oversample by a factor 100 ($f_\text{s}/f_\text{zoh}=100$) such that aliased spectra are attenuated by at least a factor 100. Note that oversampling also improves the signal-to-noise ratio.

The measurement parameters discussed above that will be used for simulations and measurements with a DST sequence are listed in Table~\ref{Table:measurementParameters}.
\begin{table}
\centering
\begin{tabular}{lcl} \hline 
Name & Parameter & Value\\
\hline
Sequence length & $N_\text{p}$ & 10002 \\
Excitation amplitude & $C$ & \SI{1}{A} (\SI{2}{A} for Fig.~\ref{fig:steadyStateNL})\\
\hline 
Zero-order hold frequency & $f_\text{zoh}$ & \SI{1.5}{kHz}\\
Zero-order hold period & $T_\text{zoh}$ & \SI{666.7}{\micro s}\\
Period length& $T_\text{p}$ & \SI{6.668}{s} \\
Lowest frequency& $f_\text{min}$& \SI{0.15}{Hz} \\
Highest frequency& $f_\text{max}$& \SI{1}{kHz} \\
\hline
Sampling frequency & $f_\text{s}$ & \SI{150}{kHz}\\
Sampling period & $T_\text{s}$ & \SI{6.667}{\micro s}\\
\hline
Number of periods & $P$ & 1\\ 
Measurement time & $T$ & \SI{6.668}{s}\\
Number of data points & $N$& 1.0002e6\\
\hline
\end{tabular}
\caption{Measurement parameters for the simulations and measurements with DST sequence.}
\label{Table:measurementParameters}
\end{table}
\section{Steady-state impedance measurements}
Batteries are nonlinear time-varying systems \cite{hallemans2023electrochemical}. However, when in steady-state and driven by a sufficiently small  zero-mean excitation, they will approximately behave as linear time-invariant systems. We exploit this to measure impedance with ternary sequences. 

Let the excitation current signal $i_\text{exc}(t)$ be a periodic ternary sequence (reconstructed with a zero-order hold) with period $T_\text{p}$ and sufficiently small amplitude $C$. The voltage response of the linear time-invariant system is \cite{hallemans2023electrochemical}
\begin{align}
    v(t)=\text{OCV}+\mathcal{F}^{-1}\{Z(\omega)I_\text{exc}(\omega)\},
    \label{eq:vtLTI}
\end{align}
with OCV the open circuit voltage of the battery, $\mathcal{F}^{-1}\{\cdot\}$ the inverse Fourier transform, $Z(\omega)$ the impedance, and $I_\text{exc}(\omega)$ the Fourier transform of $i_\text{exc}(t)$. 

\begin{figure}
\begin{center}
\includegraphics[width=\linewidth]{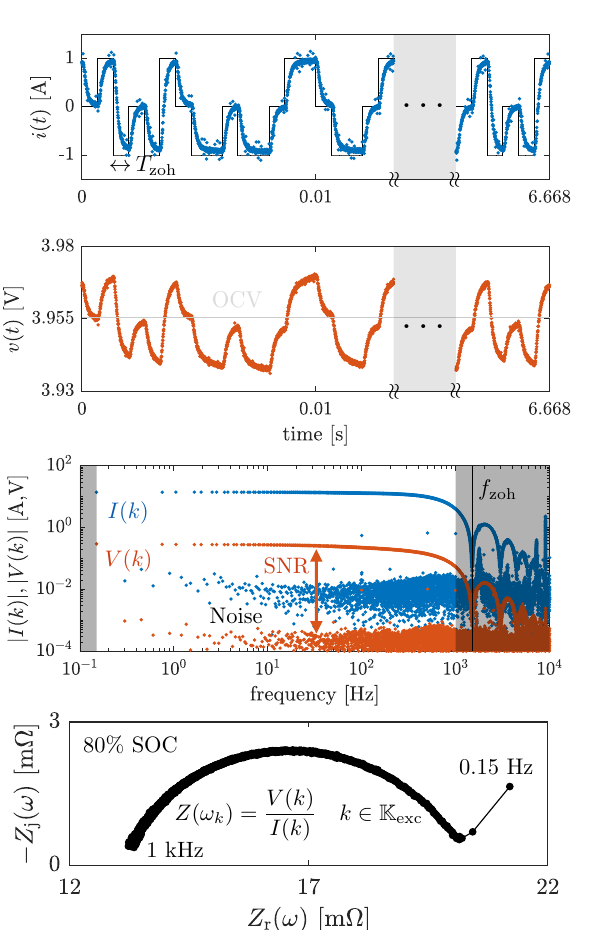} 
\caption{Measured current, voltage, and impedance data during one steady-state period (after discarding preceding period). From top to bottom: time-domain current- (1) and -voltage profiles (2), their DFT magnitudes (3), and the impedance spectrum (4).}
\label{fig:steadyStateMeasurement}
\end{center}
\end{figure}

Measuring input and output signals \eqref{eq:sampledData} as detailed in  Section~\ref{Practical measurements with PRS signals} results in data as in Fig.~\ref{fig:steadyStateMeasurement}, with DFT 
\begin{align}
    V(k)=Z(\omega_k)I_\text{exc}(k)+L(\omega_k) \qquad k>0,
\end{align}
where upper case variables refer to the DFT of the corresponding time domain data, $\omega_k=2\pi k/T$, and $L(\omega_k)$ is a transient term originating from different initial and end conditions of the finite record \cite{Schoukens2018}. This transient term can easily be removed by discarding the first few periods (e.g.\ measuring $P=2$ periods and only applying the DFT on the final one). An impedance estimate is then obtained as 
\begin{align}
    \hat Z(\omega_k)=\frac{V(k)}{I_\text{exc}(k)}\qquad k\in \mathbb{K}_\text{exc}=P\mathbb{K},
    \label{eq:ZhatLTI}
\end{align}
with $\mathbb{K}=\mathbb{K}_+\cup \mathbb{K}_-$ being the excited harmonics and $P$ the number of periods on which the DFT has been applied.

\begin{figure}
\begin{center}
\includegraphics[width=\linewidth]{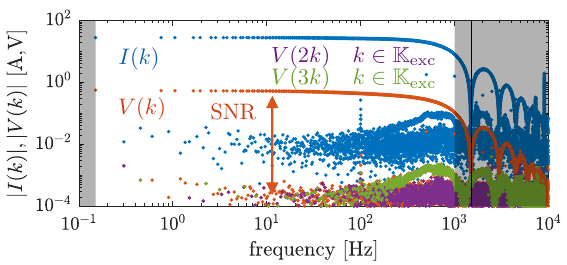} 
\caption{DFT magnitude spectra of the steady-state current and voltage measurements with highlighted 2nd- and 3rd-order nonlinearities (\SI{2}{A} excitation amplitude). 
}
\label{fig:steadyStateNL}
\end{center}
\end{figure}
Measurements also contain noise, and, hence, the signal-to-noise ratio (SNR) should be large enough to measure accurate impedance. The SNR could be improved by increasing the amplitude of the excitation $C$, but this may result in a nonlinear response---as in Fig.~\ref{fig:steadyStateNL}, where instead of only having noise at non-excited frequencies, there are also nonlinear distortions. Moreover, as the DST does not excite any second- or third-order harmonics, we can measure the level of even nonlinearities at $V(2k)$ (purple) and odd nonlinearities at $V(3k)$ (green) for $k\in \mathbb{K}_\text{exc}$. Odd nonlinearities dominate in this case. An improvement in SNR therefore comes at the cost of increased total harmonic distortion. The best linear approximation can still be obtained by dividing the voltage and current spectra at the excited harmonics \eqref{eq:ZhatLTI} \cite{schoukens2016linear}.

When performing impedance measurements in \textit{operando} conditions, one cannot remove the transient term, and there is often a superimposed drift signal too. The QRT and DST properties \eqref{eq:kposknegminusQRT} and \eqref{eq:kposknegminusDST} allow suppression of these effects, as discussed in the following section.

\section{Operando impedance measurements} \label{Operando impedance measurements}
We now consider the nonlinear time-varying system excited by an input current signal 
\begin{align}
    i(t) = i_0(t) + i_\text{exc}(t),
\end{align}
with $i_0(t)$ being a slowly varying signal to drive the system in normal operating conditions (e.g.\ constant charging current). Assuming the amplitude $C$ of $i_\text{exc}(t)$ is sufficiently small, the system will have a linear \textit{time-varying} response \cite{hallemans2023electrochemical}, 
\begin{align}
    v(t)=v_0(t)+\mathcal{F}^{-1}\{Z(\omega,t)\big(I_0(\omega)+I_\text{exc}(\omega)\big)\},
\end{align}
where $v_0(t)$ is a drift signal \cite{hallemans2022operando}, $Z(\omega,t)$ is the time-varying impedance, and  $I_0(\omega)$ is the Fourier transform of $i_0(t)$. Note that a drift signal and time-varying impedance may also be present when $i_0(t)=0$, for instance during relaxation or temperature changes.

\begin{figure}
\begin{center}
\includegraphics[width=\linewidth]{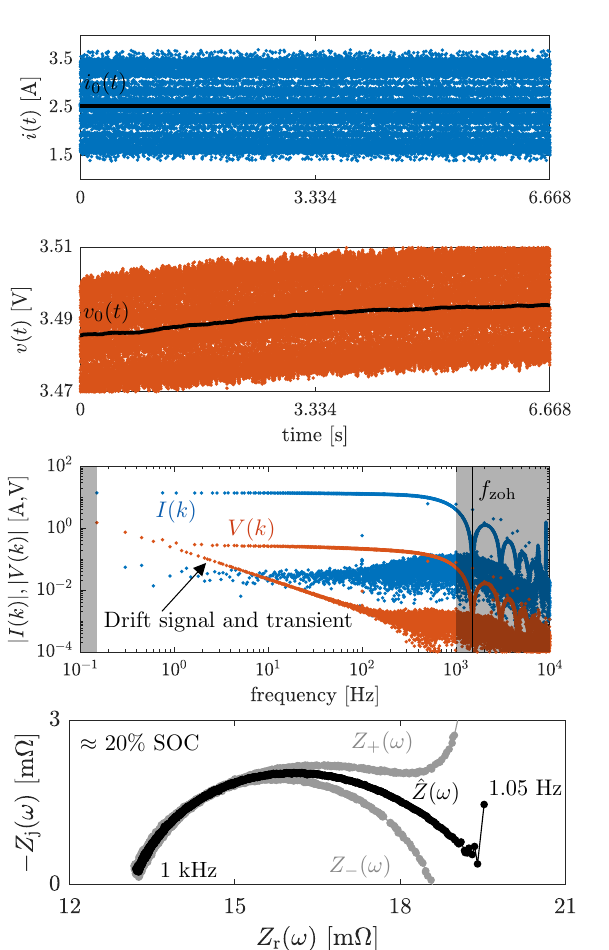} 
\caption{Measured current, voltage and impedance data during one \textit{operando} period. From top to bottom: time-domain current- (1) and -voltage profiles (2), their DFT magnitude spectra (3), and the impedance spectra (4).
}
\label{fig:LTVresponse}
\end{center}
\end{figure}

Consider an experiment over one period of the excitation signal $i_\text{exc}(t)$. When the period $T_\text{p}$ is sufficiently short, such that the system is approximately time-invariant during the applied period, the output voltage is
\begin{align}
    v(t)=v_0(t)+\mathcal{F}^{-1}\{Z(\omega)\big(I_0(\omega)+I_\text{exc}(\omega)\big)\} \qquad t\in[0,T_\text{p}],
    \label{eq:v1Poperando}
\end{align}
with $Z(\omega)$ being the average of the time-varying impedance during that period,
\begin{align}
    Z(\omega)=\frac{1}{T_\text{p}}\int_{0}^{T_\text{p}}Z(\omega,t)\mathrm{d}t.
\end{align}
Input-output data over this period was measured with a sampling period $T_\mathrm{s}$ resulting in $N_\text{pp}=T_\text{p}/T_\text{s}$ data points (see Fig.~\ref{fig:LTVresponse}). The DFT of the voltage is 
\begin{align}
    V(k) = V_0(k) + Z(\omega_k)\big(I_0(k)+I_\text{exc}(k)\big) + L(\omega_k),
    \label{eq:Yknonsteadystate}
\end{align}
where upper case variables refer to the DFT of the corresponding time domain data and $\omega_k=2\pi k/T_\text{p}$. The transient $L(\omega_k)$ is unavoidable here due to the system not being in steady-state \cite{Schoukens2018}, and due to this transient and drift signal $V_0(k)$, we do not obtain an exact measurement of the impedance $Z(\omega_k)$ by dividing $V(k)$ and $I_0(k)+I_\text{exc}(k)$ at the excited harmonics $\mathbb{K}$ \eqref{eq:ZhatLTI}. We now study removal of the effects of drifts and transients. 

\subsection{Repeated experiments for drift and transient removal}
Consider two experiments carried out under exactly the same (initial) conditions, but with two different input signals,
\begin{align}
\begin{split}
    i_+(t)&=i_0(t) + i_\text{exc}(t)\\
    i_-(t)&=i_0(t) - i_\text{exc}(t),
\end{split}
\end{align}
sampled for one period and with DFTs $I_\pm(k)$. The voltage response of these two experiments $v_\pm(t)$ is sampled for one period, and yields DFTs $V_\pm(k)$. At the excited harmonics $\mathbb{K}$ one can measure
\begin{align}
\begin{split}
        Z_\pm(\omega_k)&= \frac{V_\pm(k)}{I_\pm(k)}\quad\quad k\in \mathbb{K}\\
        &=\frac{V_0(k) + Z(\omega_k)\big(I_0(k)\pm I_\text{exc}(k)\big) + L(\omega_k)}{I_0(k)\pm I_\text{exc}(k)}.
        \label{eq:Ypm}
        \end{split}
    \end{align}
Note that the same transient term $L(\omega_k)$ is used for the two different excitations. The sign change of $i_\text{exc}(t)$, however, may lead to a (small) difference in end conditions, and, hence, a slightly different transient term. We assume that the effect of the initial conditions in the transient is dominant, and that the effect of the sign of $i_\text{exc}(t)$ on the end conditions is negligible. Rearranging these equations, the impedance $Z(\omega_k)$ can be retrieved from $Z_+(\omega_k)$,  $Z_-(\omega_k)$, $I_0(k)$, and $I_\text{exc}(k)$:
   \begin{align}
       Z(\omega_k)=&\frac{Z_+(\omega_k)+Z_-(\omega_k)}{2}+\frac{I_0(k)}{2I_\text{exc}(k)}\big(Z_+(\omega_k)-Z_-(\omega_k)\big). 
       \label{eq:Hjwkypm}
   \end{align}
Moreover, when $i_0(t)$ is a constant ($I_0(k) = 0$ for $k > 0$), \eqref{eq:Hjwkypm} is reduced to 
\begin{align}
    Z(\omega_k) = \frac{Z_+(\omega_k) + Z_-(\omega_k)}{2} \qquad k\in\mathbb{K}.
    \label{eq36}
\end{align}
This means that one can suppress the effects of the drifts and transients when performing two separate experiments with opposite sign of $i_\text{exc}(t)$, under identical (initial) conditions. This may, however, be difficult in practice. Instead, the QRT and DST sequences can be used to suppress drifts and transients from a single experiment, which is discussed next.

\subsection{Use of the QRT and DST sequences}
Property \eqref{eq:kposknegminusDST} of the DST (and QRT) sequence allows use of \eqref{eq:Hjwkypm} to suppress the drift and transient effects in a single experiment. Let an excitation $i(t)=i_0(t)+i_\text{exc}(t)$ be applied to the system with $i_\text{exc}(t)$ being a scaled QRT or DST sequence. For $k_+\in \mathbb{K}_+$ and $k_-\in \mathbb{K}_-$ being close to each other, the effect of the zero-order hold in \eqref{eq:ZOHfactor} is negligible and the DFT of the excitation satisfies 
\begin{align}
    I_\text{exc}(k_+) \approx -I_\text{exc}(k_-).
\end{align}
Alternatively, the zero-order hold effects can be taken into account for increased accuracy.
Next, let the measured current and voltage data have DFTs $I(k)$ and $V(k)$, respectively. An estimate of the impedance, with drifts and transients suppressed, can be obtained as
\begin{align}
    \hat{Z}(\omega_k) =& \frac{Z_+(\omega_k) + Z_-(\omega_k)}{2}+ \frac{I_0(k)}{2\tilde I(k)}\big(Z_+(\omega_k) - Z_-(\omega_k)\big)
\label{eq:ZreconstructionTernary}
\end{align}
for $k \in \mathbb{K}$ and with 
\begin{align}
    Z_+(\omega_k) &= 
    \begin{cases}
    \frac{V(k)}{I(k)} &\text{for } k \in \mathbb{K}_+\nonumber\\
    \text{interpolated} &\text{for } k \in \mathbb{K}_- 
    \end{cases}\\
     Z_-(\omega_k) &= 
    \begin{cases} 
    \text{interpolated} & \text{for } k \in \mathbb{K}_+\\
    \frac{V(k)}{I(k)} & \text{for } k \in \mathbb{K}_-
    \end{cases}
    \label{eq:interpolations}
    \\
    \tilde I(k) &= 
    \begin{cases} 
  I_\text{exc}(k) & \text{for } k \in \mathbb{K}_+\nonumber\\
    \text{interpolated} & \text{for } k \in \mathbb{K}_-.
    \end{cases}
\end{align}
This is similar to \eqref{eq:Hjwkypm}, however, instead of doing two experiments we take the values $Z_+(\omega_k)$ from the frequency indices $\mathbb{K}_+$ and values $Z_-(\omega_k)$ from the indices $\mathbb{K}_-$ (as the excitation contains a sign change between these indices \eqref{eq:kposknegminusDST}). Because we do not have data for $Z_\pm(\omega_k)$ at the harmonics $\mathbb{K}_\mp$ (measurements can only be taken at harmonics $\mathbb{K}_\pm$), gaps are filled using interpolation, e.g.\ linear interpolation. Fig.~\ref{fig:LTVresponse} illustrates this via the components $Z_\pm(\omega_k)$ (grey) and the reconstructed impedance $Z(\omega_k)$ (black). The simple division of spectra would have caused the impedance to jump between the two grey components, but using the technique detailed above enables suppression of transients and drift (mainly present at low frequencies). 

\subsection*{Discussion of the proposed method}
Consider what assumptions for the interpolation in \eqref{eq:interpolations} are reasonable. From \eqref{eq:Ypm}, 
\begin{align}
    Z_\pm(\omega_k)=Z(\omega_k)+\frac{V_0(k) +L(\omega_k)}{I_0(k)+ I_\text{exc}(k)}\qquad \text{for } k\in\mathbb{K}_\pm.
\end{align}
Interpolation is reasonable when the impedance $Z(\omega)$, the drift signal $V_0(k)$, the transient $L(\omega)$, slow excitation $I_0(k)$, and $I_\text{exc}(k)$ are smooth over the excited frequencies, and the distance between non-zero harmonics is sufficiently small. The input spectrum $I_\text{exc}(k)$ is smooth over the excited harmonics $\mathbb{K}_\pm$ as long as zero crossings are avoided in \eqref{eq:IZOHId}, which is the case as we only consider frequencies up to $2f_\text{zoh}/3$ while the first zero-crossing is at $f_\text{zoh}$. When the impedance is smooth, the transient is also smooth  \cite{Pintelon2012}. The arbitrary excitation $I_0(k)$ is chosen by the user, and the drift signal depends on the system. For battery experiments, the interpolation error will typically be worst at low frequencies. 

The interpolation of $Z_+(\omega_k)$ at $k\in \mathbb{K}_-$ and $Z_-(\omega_k)$ at $k \in \mathbb{K}_+$ cannot be applied for indices 
\begin{align}
    &k < \max(\min(\mathbb{K}_+),\min(\mathbb{K}_-))\quad \label{eq:lowbound}\\ 
    &k > \min(\max(\mathbb{K}_+),\max(\mathbb{K}_-)).
    \label{eq:upperbound}
\end{align}
Indeed, for these harmonics we would have to \textit{extrapolate} instead of \textit{interpolate}, which can be difficult. The high-frequency bound \eqref{eq:upperbound} is not an issue as these harmonics should anyhow be omitted from the results as these harmonics are close to $f_\text{zoh}$ and, thus, have no energy. However, at the low frequencies \eqref{eq:lowbound}, one typically loses one or two harmonics depending on the applied sequence.

In the results shown in Fig. \ref{fig:LTVresponse}, the slow signal $i_0(t)$ was a constant charging current of \SI{2.5}{A}. However, it can be any type of non-periodic signal and typically dictated by the application. Since the perturbation $i_\text{exc}(t)$ is user-defined, it can be subtracted from the measured data $i(t)$ to obtain an estimate for $i_0(t)$ and hence also $I_0(k)$ which, by default, may be unknown. 

Several signals including PRSs or multisines are suitable excitations $i_\text{exc}(t)$ for steady-state experiments. However, \textit{in operando}, more advanced frequency-domain modelling techniques are usually required to suppress drifts and transients \cite{pintelon2020frequency,hallemans2022operando}. The use of QRT or DST sequences allows suppression of these drifts and transients with a single experiment and lightweight computations \eqref{eq:ZreconstructionTernary}, making them feasible for practical use (e.g., BMS applications).

\section{Validation in simulation} \label{Simulation approach}
\begin{figure}
\begin{center}
\includegraphics[width=\linewidth]{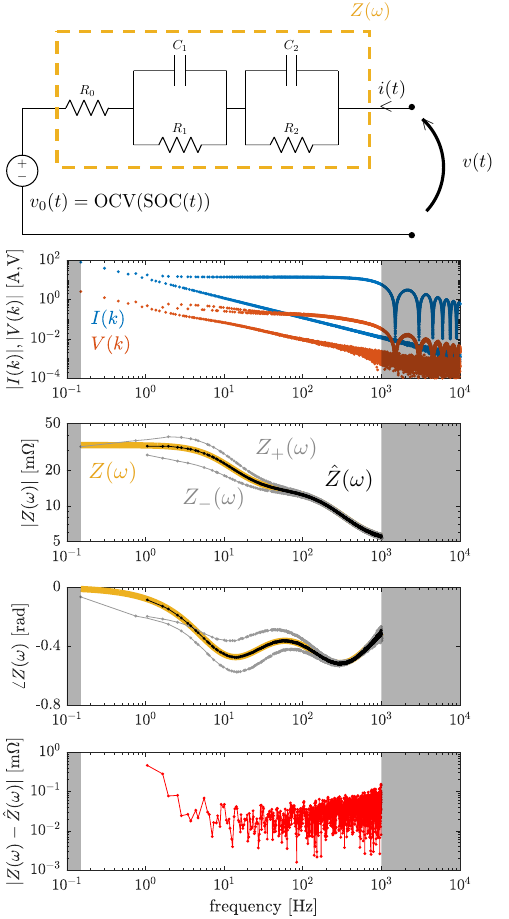} 
\caption{Results and configuration of the simulations. From top to bottom: Simulation model (1), DFT magnitude spectra of the voltage and current (2), DFT magnitude- (3) and phase spectra (4) of the impedances, and absolute error of the estimated impedance (5).}
\label{fig:SimulationError}
\end{center}
\end{figure}
We now validate the impedance reconstruction method via simulation. The model considered is an impedance superimposed on the SOC-dependent OCV, as in \eqref{eq:v1Poperando}. The impedance is a simple equivalent circuit model illustrated in Fig.~\ref{fig:SimulationError} with series resistance $R_0$ and two $RC$-branches representing the charge transfer dynamics of the two electrodes. Accordingly, the impedance is
\begin{align}
    Z(\omega)=  \frac{b_2 (j\omega)^2 + b_1 j\omega + b_0}{a_2 (j\omega)^2 + a_1 j\omega + 1}
\end{align}
with
\begin{align}
\begin{split}
    b_2 &= R_\text{s}R_1R_2C_1C_2\\
    b_1 &= R_1C_1(R_2+R_0)+R_2C_2(R_1+R_0)\\
    b_0 &=R_1+R_2+R_0\\
    a_2 &=R_1R_2C_1C_2\\
    a_1 &=R_1C_1+R_2C_2,
\end{split}
\end{align}
and component values $R_0= \SI{5}{m\Omega}$, $R_1=\SI{8}{m\Omega}$, $C_1=\SI{0.1}{F}$, $R_2= \SI{20}{m\Omega}$, and $C_2 = \SI{1}{F}$. The drift signal is obtained from OCV data
\begin{align}
    v_0(t)=\text{OCV}\big(\text{SOC}(t)\big)
\end{align}
where
\begin{align}
    \text{SOC}(t)=\text{SOC}_0+\frac{100}{3600Q}\int_0^t i(\tau)\mathrm{d}\tau,
\end{align}
with $Q=5$~Ah being the battery capacity. As a slow charging current we choose
\begin{align}
    i_0(t)=2.5-\frac{t}{2T_\text{p}}.
\end{align}
We assume the measurement parameters of Table~\ref{Table:measurementParameters}, and add zero-mean white Gaussian noise to the measured voltage and current with, respectively, standard deviations of $\sigma_v=0.5$~mV and $\sigma_i=0.5$~mA. The current and voltage spectra are shown in Fig.~\ref{fig:SimulationError}; drift signals are present in both. The true impedance $Z(\omega)$ is shown in yellow, the components $Z_\pm(\omega)$ in grey, and the reconstructed impedance $\hat Z(\omega)$ in black. The reconstructed impedance coincides with the true impedance (also notable from the error plot), which validates the proposed method. Because of \eqref{eq:lowbound}, we lose the first two excited harmonics in the reconstruction, giving a lowest frequency of \SI{1.05}{Hz} instead of \SI{0.15}{Hz}.

\section{Operando impedance measurements during fast-charging}
\begin{figure}
\begin{center}
\includegraphics[width=\linewidth]{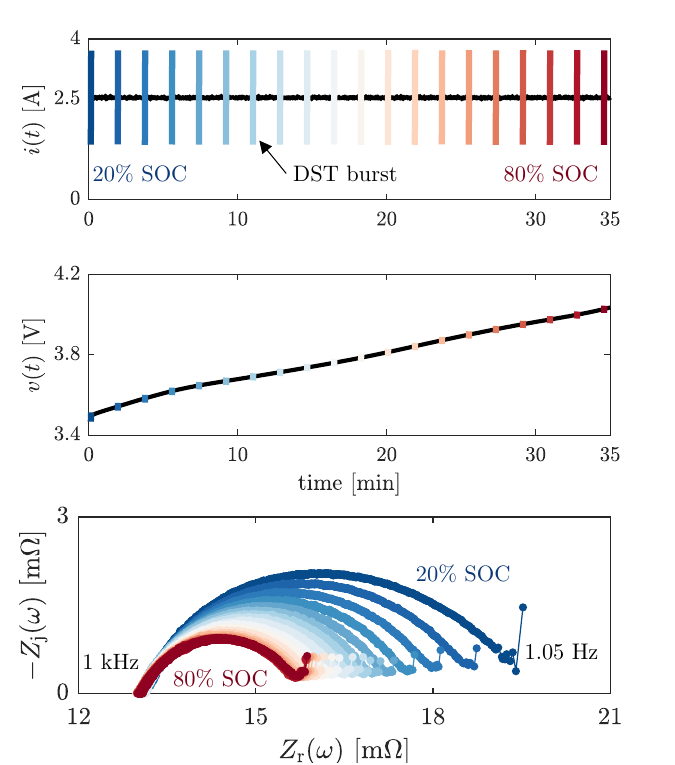}
\caption{Results of the operando impedance measurements during fast-charging across 20\% to 80\% of SOC. Top figure: current profile. Middle figure: voltage profile. Bottom figure: measured impedances.}
\label{fig:impedanceOverSOCoperando}
\end{center}
\end{figure}
We now apply the proposed techniques to measure impedance of a Li-ion battery while fast-charging. The battery is a commercial NMC 18650 cell with \SI{3.7}{V} nominal voltage, \SI{2.5}{Ah} capacity, and 0.5C of standard charging current. Initially, the cell is at 80\% SOC. The cell is connected to a programmable bi-directional power supply to apply the DST perturbation and the fast-charging current. The current and voltage are recorded with a DAQ device synchronised with the applied DST sequence. This cell and measurement setup are also used for the previous measurements (Figs.~\ref{fig:steadyStateMeasurement}, \ref{fig:steadyStateNL}, and \ref{fig:LTVresponse}). A continuous charging current of \SI{2.5}{A} (1C) was applied to the cell from 20 to 80\% SOC. The current of 1C is twice the standard charging current of the cell which was considered appropriate to replicate fast-charging conditions. A DST excitation burst with a single period of \SI{6.668}{s} was superimposed on the charging current every \SI{108}{s} to measure impedance at 20 different SOC levels (Fig.~\ref{fig:impedanceOverSOCoperando}). The DST burst and spectrum for the 20\% operating point are shown in Fig.~\ref{fig:LTVresponse}. We can see that the width of the semi-circles in the impedances decreases with increasing SOC. This is expected as the impedance depends on SOC, but also on temperature, which is increasing during fast-charging \cite{hallemans2023electrochemical}. It is noteworthy that the impedance measured in \textit{operando} provides different information compared to impedance at steady-state \cite{hallemans2022operando}. This can be noticed from these experiments, where the 80\% SOC \textit{operando} impedance has a smaller semi-circle compared to the 80\% SOC steady-state impedance of Fig.~\ref{fig:steadyStateMeasurement}. Although we have applied a constant charging current the method is also applicable for time-varying currents \cite{Sihvo2025}, making it possible to measure impedance at different operating points while fast-charging an electric vehicle. The proposed PRS methods are also applicable during relaxation, and arbitrary application use profiles (e.g.\ during EV driving  \cite{Sihvo2025}, or energy trading for a stationary battery).


\section{Conclusions} \label{Conclusions}
\textit{Operando} impedance measurements are useful for monitoring batteries in the field. In this work, we presented pseudo-random sequences that can be applied with low-cost electronics, allowing measurement of impedance data in operational conditions with computationally efficient data processing. We discussed the QRT and DST sequences, which are eigenvectors of the DFT matrix at the excited harmonics. This property makes the drifts- and transients-affected impedance spectrum to separate into two components from which the underlying, true impedance can easily be extracted. 

This technique is promising for economically feasible impedance measurements within a BMS during fast-charging, relaxation, or even driving. Short bursts of the proposed DST sequence can be applied over a charging current to extract \textit{operando} impedance, with applications ranging from state-of-health prognostics to parametrizing models for fast-charging.

\section*{Appendix: Proof of \eqref{eq:Theorem1Eigenvector}}
Recall the following property of a repeated signal. Let $u(n)$ be a discrete signal with length $N$ and DFT $U(k)$. Assume $u(n)$ is repeated $P$ times, resulting in $u_P(n)$. The DFT of $u_P(n)$ is
\begin{align} 
    U_P(k) = \begin{cases}
        \sqrt{P}U\big(\frac{k}{P}\big) & \text{for } k=0,P,2P,...,(N-1)P\\
        0   &\text{else}.
    \end{cases}
    \label{eq:spectrumRepeated}
\end{align} 

The vector $\mathbf{a}$ is constructed by repeating the special sequence $\mathbf{u}_\text{special}=[0\ -1\ -1\ 0\ 1\ 1]$ \eqref{eq:uspecial} (with DFT $\mathbf{U}_\text{special}=[0\ j\sqrt{2}\ 0\ 0\ 0\ -j\sqrt{2}]$) $N_\text{QRT}$ times. Using \eqref{eq:spectrumRepeated}, we find that
\begin{align}
A(k) = 
  \begin{cases} 
	j\sqrt{2N_\text{QRT}} & \text{for } k =N_\text{QRT}\\
    -j\sqrt{2N_\text{QRT}} & \text{for } k =5N_\text{QRT}\\
	0  &\text{else}.
 \end{cases}
 \label{eq:Ak}
 \end{align}

The vector $\mathbf{b}$ is generated by repeating a QRT sequence $\mathbf{u}_\text{QRT}$ 6 times, hence its DFT is
\begin{align}
B(k) = 
 \begin{cases} 
	\sqrt{6}U_\text{QRT}\big(\frac{k}{6}\big)  & \text{for } k=6,12,\hdots,6(N_\text{QRT}-1)\\
	0   &\text{else.}
 \end{cases}
 \label{eq:BkTheorem1}
\end{align}
Using the fact that the QRT sequence is a completely multiplicative function \cite{Pei2008}, 
\begin{align}
    u_\text{QRT}(n) = \frac{u_\text{QRT}(\left[dn\right]_\text{mod $N_\text{QRT}$} )}{u_\text{QRT}(d)} \hspace{2mm} \text{for } 0\leq d,n < N_\text{QRT},
\label{eq:multiplicativeFunction}
\end{align}
with $u_\text{QRT}(6)=r$ ($r\in\{-1,1\}$ depending on $N_\text{QRT}$), we find that for $n=0,1,...,N_\text{QRT}-1$ 
\begin{align}
    b(6n)&=u_\text{QRT}([6n]_{\text{mod} N_\text{QRT}})=r u_\text{QRT}(n)\nonumber\\
    &=\frac{r}{\lambda_\text{QRT}}U_\text{QRT}(n)=\frac{r}{\sqrt{6}\lambda_\text{QRT}}B(6n).
\end{align}
Hence,
\begin{align}
    B(k)=
    \begin{cases} 
	r\sqrt{6}\lambda_\text{QRT}b(k)&\text{for } k=0,6,12,\hdots,6(N_\text{QRT}-1),\\
	0   &\text{else}.
 \end{cases}
 \label{eq:eigenvectorb}
\end{align}
Therefore $\mathbf{b}$ is an eigenvector of the DFT at its excited harmonics.

The effect of the element-wise multiplication of $a(n)$ and $b(n)$ in the time-domain  \eqref{eq:uDSTn} corresponds to circular convolution of $A(k)$ and $B(k)$ in the frequency-domain,
\begin{align} 
    U_\text{DST}(k) = \frac{1}{\sqrt{N_\text{DST}}}\sum_{m=0}^{N_\text{DST}-1}A(m)B([k - m]_\text{mod $N_\text{DST}$}).
\label{eq:convolution}
\end{align}
$A(k)$ has only two non-zero values \eqref{eq:Ak}, hence, we have that
\begin{align}
    U_\text{DST}(k) = \frac{j}{\sqrt{3}}\big(&B([k - N_\text{QRT}]_\text{mod $N_\text{DST}$})\nonumber\\
    &-B([k - 5N_\text{QRT}]_\text{mod $N_\text{DST}$})\big).
\end{align}
$B(k)$ \eqref{eq:eigenvectorb} is only non-zero for $k\in \{6,12,\hdots,6(N_\text{QRT}-1)\}$ \eqref{eq:BkTheorem1}, hence, the term $B([k - \alpha N_\text{QRT}]_\text{mod $N_\text{DST}$})$ (with $\alpha=1$ or $5$) is only non-zero when
\begin{align}
    &k-\alpha N_\text{QRT}=6p &\text{for }k\geq \alpha N_\text{QRT}\\
    &k-\alpha N_\text{QRT}+N_\text{DST}=6p &\text{for }k<\alpha N_\text{QRT},
\end{align}
with $p=1,2,\hdots,N_\text{QRT}-1$. Now, for $N_\text{QRT}=7 + 6q$ ($q \in \mathbb{N}$) \eqref{eq:NDST}, $B([k - \alpha N_\text{QRT}]_\text{mod $N_\text{DST}$})\neq0$ for $k\in\mathbb{K}_\alpha$ with
\begin{align}
    &\mathbb{K}_1=\{1+6p\ \vert \ p=0,1,\hdots,N_\text{QRT}-1\}\setminus N_\text{QRT}\\
    &\mathbb{K}_5=\{5+6p\ \vert \ p=0,1,\hdots,N_\text{QRT}-1\}\setminus 5 N_\text{QRT}
    .
\end{align}
The non-zero values of $U_\text{DST}(k)$, hence, occur at the harmonics 
\begin{align}
      k\in\mathbb{K}=&\{1+6p,5+6p \ | \ p=0,1,\hdots,N_\text{QRT}-1\}\nonumber\\
      &\setminus \{N_\text{QRT}, 5N_\text{QRT}\}.
      \label{eq:Theorem1_I_appendix}
\end{align}
Using the eigenvector property of $b$ \eqref{eq:eigenvectorb}, and noting that $b([n - \alpha N_\text{QRT}]_\text{mod $N_\text{DST}$})=b(n)$ (for both $\alpha=1$ and 5) due to the periodicity over $N_\text{QRT}$ samples and the fact that $N_\text{DST}=6N_\text{QRT}$, we find that
\begin{align}
    U_\text{DST}(k)=  \begin{cases} 
	j\sqrt{2}r\lambda_\text{QRT}b(k) & \text{for } k \in \mathbb{K}_1\\
        -j\sqrt{2}r\lambda_\text{QRT}b(k) & \text{for } k \in \mathbb{K}_5\\
	0  &\text{else}.
 \end{cases}
\end{align}
Now, as $a(n)$ is periodic over 6 and we know its values, we find that
\begin{align}
    b(n)=\frac{u_\text{DST}(n)}{a(n)}=\begin{cases} 
	u_\text{DST}(n) & \text{for } n\in \mathbb{K}_1\\
        -u_\text{DST}(n) & \text{for } n\in \mathbb{K}_5\\
	0  &\text{else},
 \end{cases}
\end{align}
hence,
\begin{align}
    U_\text{DST}(k)=  \begin{cases} 
	j\sqrt{2}r\lambda_\text{QRT}u_\text{DST}(k) & \text{for } k\in\mathbb{K}\\
	0  &\text{else}.
 \end{cases}
\end{align}
This is also valid for the other choice $N_\text{QRT}=5 + 6q$ \eqref{eq:NDST}.

\bibliographystyle{IEEEtran}        
\bibliography{main}           

\end{document}